\documentclass[5p,fleqn]{elsarticle} % 5p is the 2‑column journal look

\usepackage{graphicx}
\usepackage{changepage}
\usepackage{amsfonts}
\usepackage{tabularx} % in preamble
\usepackage{amsmath}  % make sure this is in your preamble
\usepackage[version=4]{mhchem}  % Optional, for cleaner isotope and chemical formatting
\usepackage{hyperref}
\usepackage{float}
\usepackage{xurl}
\usepackage{booktabs}
\usepackage{xcolor}

\begin{document}
\begin{frontmatter}

\title{Updated Simulation of GRETA Detector Response and Exploration of Temperature Sensitivity}
\author[1,2]{Arin Manohar}
\author[1]{Mario Cromaz}
\author[1]{Christopher Campbell}
\author[1]{Heather Crawford}
\author[1]{Marco Salathe}
\address[1]{Nuclear Science Division, Lawrence Berkeley National Laboratory, Berkeley, CA 94720 USA}
\address[2]{Department of Physics, University of California, Berkeley, CA 94720, USA}

\begin{abstract}
The Gamma-Ray Energy Tracking Array (GRETA) is a next-generation gamma-ray spectrometer designed to push the frontiers of nuclear structure and astrophysics experiment.  Its high sensitivity is enabled by high-precision localization of gamma-ray interactions within its active detector volume, and the subsequent tracking of gamma-ray scattering sequences.  In order to perform gamma-ray tracking, we need to simulate signal generation in the detectors accurately.  This requires both accurate calculations of charge movement in the semiconductor volume, as well as a faithful reproduction of real-world experimental effects such as the electronics response.  This work addresses the fidelity of the calculated signals for GRETA in two ways.  An updated approach has been applied to find an optimized parameterization of the electronics response, while the impact of the detector temperature was also explored to best reproduce experimental signals and improve the position localization performance for GRETA.  The results suggest that the electronics response can be simplified without impacting performance, and that the response correction parameters can effectively compensate for the changes in signal which arise due to the crystal temperature, resulting in minimal sensitivity of the position resolution to the assumed temperature.
\end{abstract}


\end{frontmatter}

\section{Introduction}

The Gamma-Ray Energy Tracking Array (GRETA)~\cite{gretaurl} is the realization of a nearly $4\pi$ gamma-ray tracking detector, which will enable sensitive measurements in nuclear structure and reactions and nuclear astrophysics research.  GRETA builds on the Gamma-Ray Energy Tracking In-beam Nuclear Array (GRETINA)~\cite{GRETINAPaschalis}, using the same Quad Detector modules, and the same overall signal processing approach.  However, GRETA employs a total of 120 large-volume high-purity Germanium (HPGe) crystals, arranged in a close-packed configuration to cover $\sim$80\% of the solid angle surrounding a target, considerably increasing the efficiency for gamma-ray detection as compared to GRETINA.  

The geometry and principles of operation for the GRETA$\slash$GRETINA detectors have been described in earlier publications~\cite{GRETINAPaschalis}.  Briefly, each crystal is 9~cm in length, with a 8~cm diameter cylindrical shape at the rear, which tapers to one of two irregular hexagonal front face geometries (A- and B-type crystals).   Two crystals of each type are housed together to form Quad modules, which close-pack with 30 Quad modules combining with 12 pentagonal 'holes' to cover the solid angle surrounding the target position.  An inner coaxial contact provides the full-volume readout for each crystal, while the outer contact of each HPGe crystal is segmented 6 ways longitudinally and laterally producing 36 voxels, which are read out individually as 36 segment contacts in addition to the core.  Gamma-rays which interact in a given crystal deposit energy in one or more of these voxels, generating a set of distinct segment + core contact current signals based on net charge collection as well as signals induced on segments neighboring those with net charge.  Through detailed simulations, the signals generated on these contacts are calculated by considering how a unit test charge deposited at a particular position inside the crystal will drift.  For GRETA$\slash$GRETINA, this is done for  $\sim$231,000 points inside the crystal on a fixed grid with an average 1mm spacing between neighboring grid positions. The set of all of these points and their corresponding signals is called a detector basis. Using such a basis, we can construct a realistic signal for multiple interaction points via a linear combination of basis signals subject to some constraints. In experiment, the gamma-ray interaction points are located, and their relative energy deposition determined, by comparing linear combinations of these basis signals against the set of observed experimental signals.  This process of signal decomposition has enabled reconstruction of gamma-ray interactions with $\sigma$ better than 2~mm resolution~\cite{GRETINAPaschalis, BruyneelADL, DirkGRETINAPerformance}.

However, the accuracy of this interaction-point localization fundamentally depends on the fidelity of the calculated basis.  The simulations for each detector are complex and depend on numerous parameters.  The simulations are divided into two steps.  The first step models the charge transport in the crystal, simulating the signals generated on the segment pads by drifting charge carriers in the calculated electric field of the detector. This relies on several inputs including the crystal geometry and parameterizations for the mobilities of electrons and holes, which depend on impurity concentrations and temperature.  From these inputs, electric fields and weighting potentials are calculated, which enables the signal simulations.   	

While the first step in the process provides what is known as a ``pristine basis'', to compare with measured experimental signals, a second phase of the calculation is required to account for the electronic response of the detector readout system. Due to close packing of segment channels and properties of the preamplifier inputs, both integral and differential crosstalk arise between segments. In addition, the preamplifiers shape the signals and this shaping differs slightly channel by channel. In addition, there can be relative timing delays between different signal chains and global timing delays that offset all signals. All of these effects modify the pristine signals generated in the first phase and must be modeled and corrected before the ``corrected basis'' can be reliably used to with experimental data for position reconstruction.

In this work, two aspects of the production of the corrected signal bases used in GRETA$\slash$GRETINA were explored.  First, we simplified and improved the parameterization of the electronics response function used, as well as the code which is used to fit the response parameters.  We confirmed the quality of the signal decomposition in terms of the position resolution, using a basis created with the simplified response model.  Second, we explored the sensitivity of the results of signal decomposition to the crystal temperature assumed in the first step of basis production.  The position resolution from experimental pencil beam data does show some sensitivity to the assumed temperature, but initial results suggest that a more thorough investigation, and perhaps a more complex treatment of temperature may be required for fully optimized performance.

\section{Basis Generation}
\label{sec:basisGen}

The calculation of a signal basis for a given GRETA$\slash$GRETINA crystal follows a now well-established procedure, which has been discussed in previous publications (e.g.~\cite{Prasher2017}), and used for the detectors of the GRETINA array.  This procedure is briefly summarized in the following. 

\begin{enumerate}
\item First, the electric field within the HPGe volume, and the weighting potential associated with each electrical contact are calculated using the \texttt{fieldgen}~\cite{radford2018} software code.  The unique crystal properties, including geometry, vendor-provided impurity profiles, depletion voltage and temperature are provided as inputs.  Following a scaling of the impurity profile to reproduce the depletion voltage measured for the crystal, the electric potential, electric field, and weighting potentials are solved for on a 1~mm grid using an iterative relaxation approach.  
\item With the electric field and weighting potentials determined, the signals associated with drift of a unit charge from a given position in the crystal can be calculated using the \texttt{siggen}~\cite{radford2018} code.  The electric field is used to to calculate the drift path of both electrons and holes from a given starting position within the crystal volume, with the drift velocity given by the combination of the electric field and charge carrier mobility, $\mu$.  This is discussed further in Section~\ref{sec:tempDep}.  Once the drift paths are established, the signals on each electrical contact are based on the Schokley-Ramo theorem~\cite{Schockley, Ramo} which relates moving charge to induced current.  By repeating the calculation of drift path and signals for point charges deposited on a grid of initial positions, a raw, or pristine, basis is generated.
\item Once a pristine basis has been generated, the next step is to establish the electronics response correction, which accounts for the modification of the raw basis signals as a result of the detector readout system.  For this, the ``superpulse'' method is used, in which sets of averaged signals are generated both experimentally (based on a flood-field measurement with $^{60}$Co) and from the pristine basis, and fit against each other to find an optimal set of parameters for the chosen electronics response function.  This fit has been performed to this point using non-linear least-squares minimization and a parametrized electronics response function.  More details on the superpulse method, and the updates to the fitting approach and electronics response function are described in Sections~\ref{sec:SPmethod}.
\item The final step in the production of the calculated basis for a crystal is the application of the electronics response to all of the signals in the pristine basis.  The same response function parametrized by the corresponding optimal electronics parameters is applied to the signals of each basis point in the pristine basis, using the set of parameters determined in the superpulse method.  The corrected basis is used in signal decomposition for comparison to experimental pulses event-by-event to localize gamma-ray interaction positions.
\end{enumerate}

\section{Superpulse Method and Electronic Response Parameterization}
\label{sec:SPmethod}

\subsection{Experimental and Simulated Superpulse Production}

The ``superpulse'' method for constraining the electronics response is based on the comparison between experimental superpulse data and the equivalent simulated and calculated detector response, where a ``superpulse'' is defined as the set of averaged signal traces (all 36 segments and the core contact) for events corresponding to a net charge collected on a single crystal segment.  

\begin{figure*}[htbp]
    \centering
    \includegraphics[height=0.4\textheight]{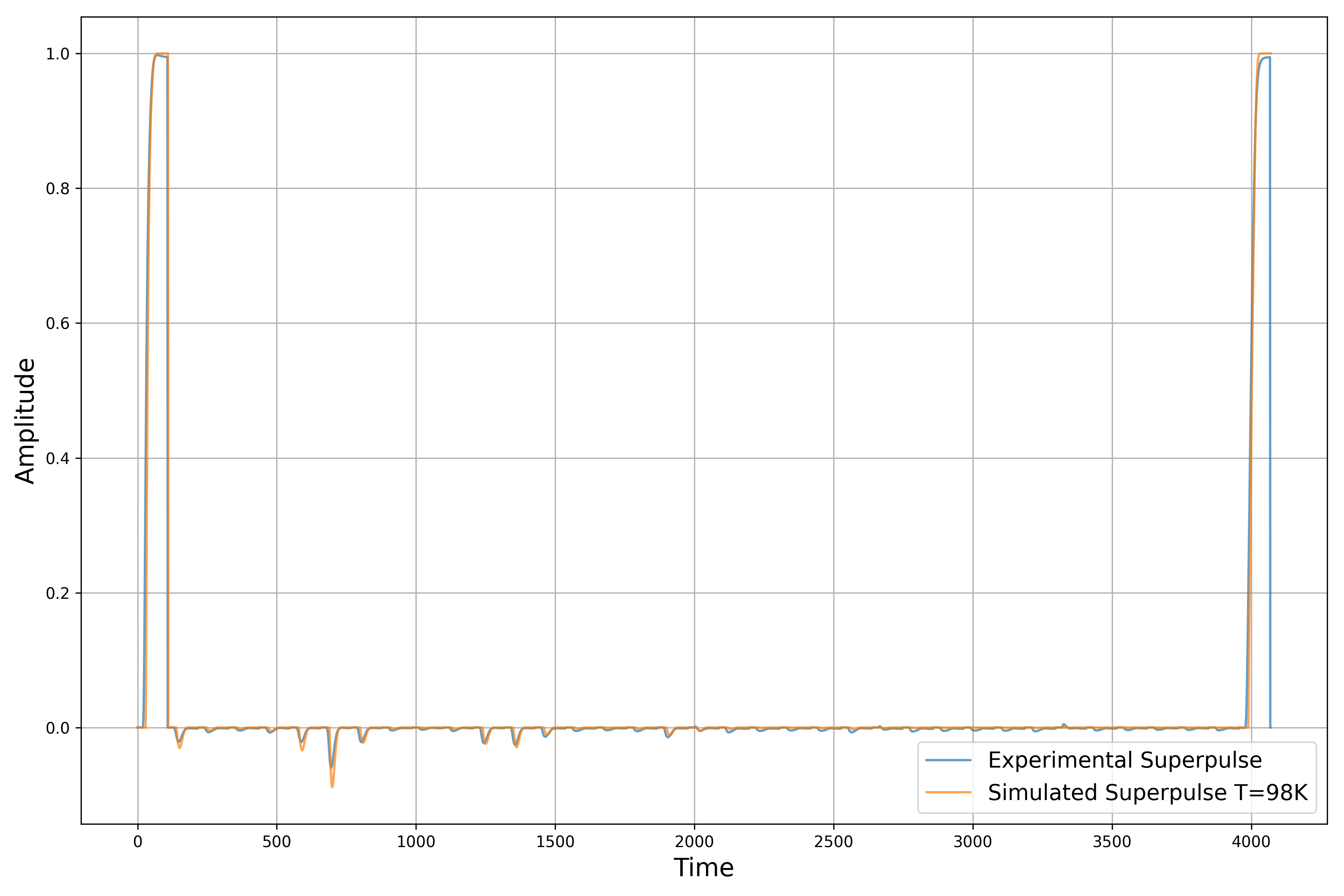}
    \vspace{-1em}
    \caption{Example of a superpulse (both experimental and simulated) concatenated pulse for events in which segment 0 (first segment on the left) records a net charge.  The full superpulse is the set of these concatenated pulses for each segment recording a net charge. }
    \label{fig:superpulse}
    \vspace{-1em}
\end{figure*}

Experimentally, superpulses are typically generated from flood-field measurements using a $^{60}$Co source.  Events are selected which satisfy the requirement of a single segment recording a net energy corresponding to 1332~keV, with the core-contact also recording this energy.  All events corresponding to full energy depositions in a given segment are averaged, meaning that the signals from all 36 segments and the core-contact are concatenated and an average concatenated pulse determined.  This is done for full energy depositions in each segment, yielding a set of 36 average concatenated pulses, the experimental superpulse.  Figure~\ref{fig:superpulse} shows an example of the concatenated pulses for events where segment 0 (the first segment on the left) records a net charge.  Experimental superpulses are produced by standard analysis codes used for GRETA$\slash$GRETINA datasets.

The simulated superpulse is similarly produced.  A GEANT4 simulation of the same geometry as the flood-field measurement, and using the same source energy is performed to generate sets of interaction points/energy depositions within the crystal volume.  Events are filtered to identify those which satisfy the condition of all charge corresponding to a 1332~keV gamma-ray being collected within a single segment volume.  For each event which satisfies this condition, the basis point in the pristine basis closest to the energy deposition location is identified and the corresponding signals are linearly combined with those from other energy depositions in the same event, weighted by the deposited energy fractions.  Events signals constructed in this way are then averaged in the same way as is done for the experimental superpulse, yielding a simulated superpulse in the same format as the experimental reference.  

In the basis production pipeline, the simulated superpulse was produced as part of the C-based basis production software.   A new implementation of the simulated superpulse production has been developed in Python, leveraging the K-D Tree data structure \cite{bentley1975kdtree} and pandas~\cite{pandas} to allow for efficient nearest neighbor lookup and gating condition application. 

%\heather{Arin - can you add something about the work you did on this? Doesn't need to be long, just a sentence is fine.}
%\arin{The interaction positions and energies were loaded into a Pandas Dataframe allowing for efficient nearest neighbor lookups and gating condition enforcement. }

\subsection{Electronics Response Function Parameterization}
\label{subsec:electronicsResponse}

To capture the impact of the readout electronics for GRETA$\slash$GRETINA, an electronics response function was developed with a total of 669 parameters, accounting for signal shaping, differential cross-talk (DXT), integral cross-talk (IXT) and signal time delays/shifts.  The parameterization is outlined in the following.  

The first transformation applied accounts for the signal shaping of the preamplifier electronics.  This is modeled as a convolution of the pristine signal with an exponential function to account for the shaping introduced by the limited frequency response of the preamplifier.  Each segment and the central contact has a unique shaping time constant $\tau$ associated with it, giving rise to a total of 37 parameters associated with signal shaping.  Two additional times account for segment-segment and segment-core cross-talk with respect to shaping, which apply the an exponential shaping to the derivatives of the traces. 

After shaping, DXT is considered in the electronics response.  Differential crosstalk arises due to coupling between closely spaced signal paths within the detector, and is modeled as a coupling between the time-derivatives of signals on neighboring channels.  Differential cross-talk coefficients connect pairs of segments, and are assumed to be significant for signals which are close in the electronics layouts.  As DXT is assumed to be symmetric aside from the core-contact, taking 17 neighbors as important for a given segment,  342 parameters are associated with DXT. 

IXT models capacitive coupling between crystal segments, in which a portion of the signal in one segment results in a response in neighboring segments.  This coupling is assumed to be non-negligible in segments which are physically close, and is modeled as a coupling between the integral signals on these segments.  IXT is not assumed to be symmetric for segment pairs.  Assuming five important neighboring segments for each segment, and integral cross-talk of all segments with the core contact, a total of 216 IXT parameters are included in the electronics response function.  

Finally, in the original electronics response function, two time delays per segment were included to account for channel-to-channel time offsets in the electronics, resulting in an additional 72 parameters to ensure time alignment.  

The electronics response described above incorporates a total of 669 free parameters to be constrained by the superpulse fit procedure.  This parameterization has been the function in use for GRETINA for the past 10+ years, without modification.  However, this parameter set includes values which remained in use due only to operational legacy.  Specifically, the now robust time synchronization of the GRETINA systems essentially negates the need for 72 individual time delays, and a single global delay is expected to provide the required delay functionality in the electronics response.  In addition, the shaping applied to the signal derivatives is no longer relevant.  As such, a total of 73 parameters can be removed based on the physics of the system, reducing the electronics response to a total of 596 free parameters.  

%\heather{We need to somehow link the new implementation of the SP fitting etc. to the change in the proposed response.  Was there a sensitivity study or a correlation analysis which suggested consolidation of the delay parameters?  Or was this an informed guess?  Then we should summarize the changes to the response function.}  \arin{Mario can explain more, but the 36 delay0 and 36 delay1 were originally chosen due to the old electronics design of GRETA. Since then, the electronics have simplified, and there is now reason to only include a global delay as opposed to delays per segment and per signal. }

%In total, we have $37$ shaping time parameters $\tau_i$ (one per signal trace), $342$ DXT coefficients $D_{ij}$ over the prespecified set of differential crosstalk pairs $\mathcal{S}^{\text{(DXT)}}$, $216$ IXT coefficients $I_{ij}$ over the prespecified set of integral crosstalk pairs $\mathcal{S}^{\text{(IXT)}}$, and a single global time offset parameter $\Delta t$, summing to $596$ parameters.

\subsection{Superpulse Fitting and Corrected Basis Production}
\label{subsec   :optimal_params}

The total electronics response function can be modeled as a parameterized transformation $f(\hat{s}, \vec{\theta})$ which acts on the input superpulse $\hat{s}$, where $\vec{\theta} = (\theta_1, \theta_2, \dots, \theta_{596})$ represents the parameters of the response function, namely all the shaping time constants, DXT and IXT coefficients, and the global delay. This was similarly true for the previous parameterization which included individual channel delays, simply with additional parameters.  

The original implementation of the superpulse fitting was written in C, and used the Levenberg-Marquardt, or damped least-squares, optimization to fit the optimal response parameters~\cite{bevington}.  As part of the update to the basis production pipeline, this fitting step has also been re-factored in Python.  

To determine the optimal parameter vector $\vec{\theta}^*$, we minimize the mean squared error (MSE) loss between the transformed simulated superpulse and the experimentally measured one:
\begin{equation}
\begin{aligned}
\vec{\theta}^* &= \arg \min_{\vec{\theta}} \, L\big(f(\mathbf{x}_{\text{sim}}; \vec{\theta}), \, \mathbf{x}_{\text{exp}}\big) \\
&= \left\|f(\mathbf{x}_{\text{sim}}; \vec{\theta}) - \mathbf{x}_{\text{exp}}\right\|^2
\end{aligned}
\end{equation}

%\heather{Contrast the new implementation with the original C code in terms of the SP fit -- what have we gained? If the text below if accurate, we need to add citations for "trust-region reflective algorithm" etc. to back up the claims about convergence and efficiency that follow.}
The new nonlinear least-squares optimization is carried out using a trust-region reflective algorithm \cite{trf}, which balances convergence stability and computational efficiency, even in the presence of sparse or unavailable gradient information. Upon convergence, we obtain both the optimal parameter set $\vec{\theta}^*$ and the corresponding minimized loss value.  

%\heather{How is convergence checked in the new implementation?} \arin{Convergence is reached when any of the following residual tolerance conditions occur for optimization iteration $k$:   $||\vec{\theta}_k - \vec{\theta}_{k-1}||_2 \le ||\Delta\vec{\theta}_{tol}||_2$  or  $||f(x;\vec{\theta}_k) - f(x;\vec{\theta}_{k-1})||_2 \le ||\Delta f(x;\vec{\theta})_{tol}||_2$ or $||\nabla f(x;\vec{\theta}_k) - \nabla f(x;\vec{\theta}_{k-1})||_2 \le ||\Delta( \nabla f(x;\vec{\theta})_{tol})||_2$}

Once the optimal electronics response function parameters are determined, the response is applied to the set of signals for each basis point in the pristine basis, to generate the final corrected basis which is used in experiments for localizing gamma-ray interaction positions, in a process known as signal decomposition\cite{radford2018, cromaz2024signaldecomp}.  The complexities of this algorithm are not the focus of this work, but at a high level the signal decomposition finds the best match to the set of signals from each experimental event by fitting the experimental pulse with a linear combination of basis signals.  As such, the fidelity of the corrected signal basis with respect to experimental data is critical.

%\heather{Stopping here for tonight  (12/2/25) -- I will keep working on the remaining.  But one thing we need to have is a comparison of the performance for a pencil beam of the original electronics response vs. the new, if we want to highlight this as a result}.

\subsection{Performance of Simplified Electronic Response Function}
\label{subsec:pencilBeamAndResponse}

As discussed in Section~\ref{subsec:electronicsResponse}, with the updated implementation of the simulated superpulse production and superpulse fitting, the electronics response function for GRETA$\slash$GRETINA was simplified, removing extra shaping parameters and consolidating the 72 delay parameters to a single global delay parameter.  While these modifications were physically motivated, it is still important to confirm that the performance of signal decomposition is not negatively impacted with these changes.  To this end, the position reconstruction performance was investigated by considering the resolution obtained for a pencil-beam data set.  

\begin{figure*}[htbp]
    \centering
    \includegraphics[height=0.25\textheight]{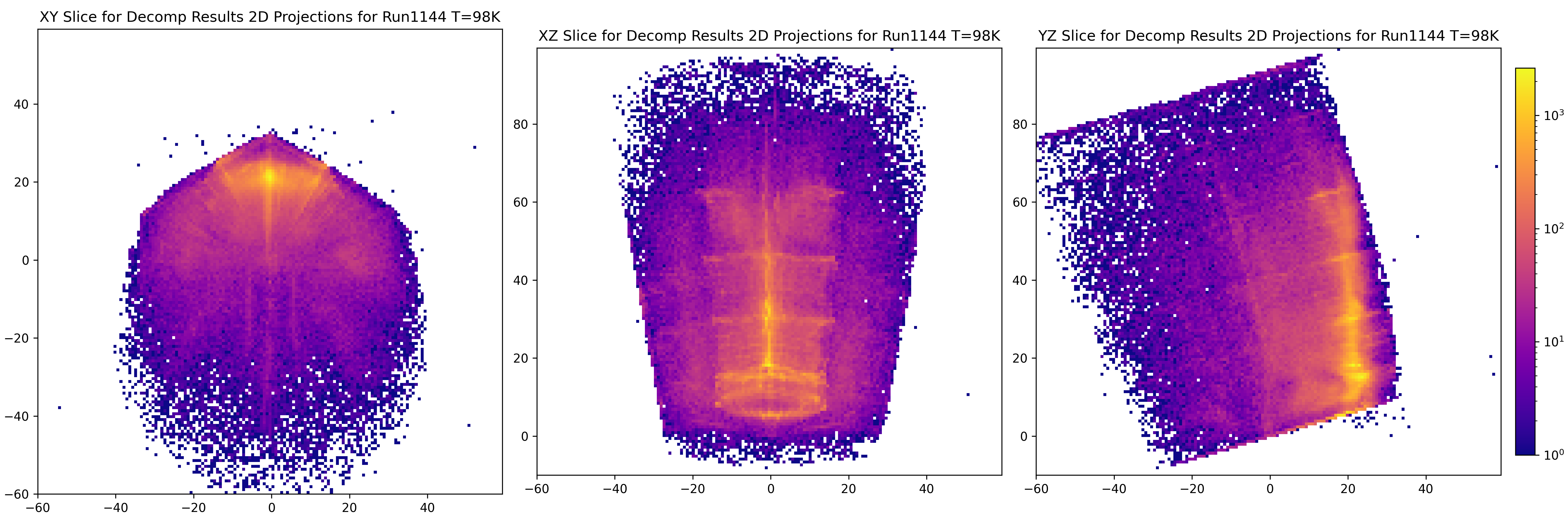}
    \vspace{-1em}
    \caption{Localized interaction positions for a typical pencil beam measurement, plotted in a series of 2D projected histograms after rotation of the intrinsic crystal coordinates to account for the tilt of the crystal during measurement.  The leftmost figure shows a projection in (x', y'), the center figure in (x', z') and the right-most figure in (y', z').  Brighter regions indicate more interactions occurring at those positions.  This figure shows all interaction positions, with no event selection cuts.}
    \label{fig:pencilBeam}
    \vspace{-1em}
\end{figure*}

Pencil beams are measurements made with a highly-collimated ``beam'' of gamma rays directed through a mechanical collimator, typically from a strong $^{137}$Cs source, producing first-hit energy depositions localized along a straight line within the detector.  An example of a pencil beam data set is shown in Figure~\ref{fig:pencilBeam}.  By selecting events with full-energy depositions and projecting the pencil beam data in the rotated coordinate frame, it is possible to extract effective position resolutions based on the distribution of interactions in a pencil beam data set.

\begin{figure}[htbp]
    \centering
    \includegraphics[width=\linewidth]{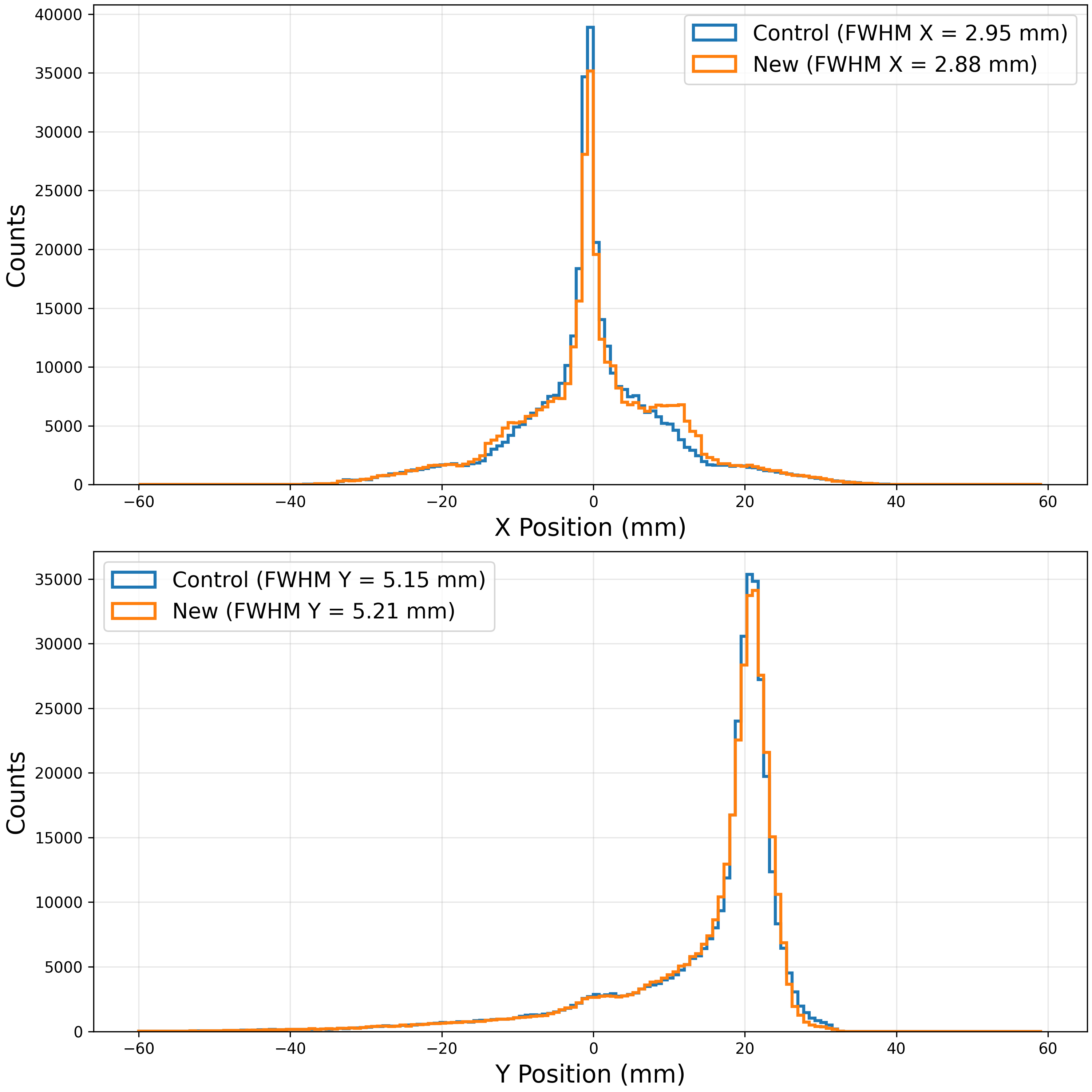}
    \vspace{-1em}
    \caption{Top panel: x' projection of pencil beam for crystal A35 after signal decomposition using the original crystal basis with 669 electronics response parameters (orange) and the reduced parameter set (blue).  Bottom panel: The same as above, but for the y' projection. }
    \label{fig:xy}
    \vspace{-1em}
\end{figure}

A pencil beam data set taken using the 662~keV transition in $^{137}$Cs in GRETA crystal A35 (an A-type crystal which was housed as position 2 of Q18) \cite{riley2020ucgretina} was analyzed through signal decomposition using a corrected basis produced in full with the original basis production pipeline and 669 parameters electronics response correction (Basis-669), as well with as basis produced with the updated implementation of the simulated superpulse preparation and fitting, using the reduced electronics response with 596 parameters (Basis-596).  The extracted position resolutions, taken as the numerical full-width at half-maximum of the distributions in x' and y' (shown in Fig.~\ref{fig:xy}) are 2.95~mm, 5.15~mm and 2.88~mm and 5.21~mm for Basis-669 and Basis-596 respectively. % The goodness-of-fit metrics for the superpulse fitting are similarly comparable \heather{Arin, can you confirm?}.

%\heather{Arin - can you add the information for the control pencil and your new basis with the reduced parameter set?  Maybe it would also be good to show one example of an x' and y' distribution at this point in the paper.  We can then remove the one in Fig. 6 and focus just on the FWHM in the temperature discussion.}\arin{Added x y hists in figs control vs new decomp hists A35}

\section{Sensitivity of Position Resolution to Temperature}

While the electronics response function is critical to producing a corrected basis with high fidelity, there are additional parameters which impact the faithfulness of the basis, related to the crystal parameters.  Specifically, the detector impurity profile, depletion voltage and temperature significantly impact the final crystal basis.  These are input in the first stage of the basis production, at the point of calculation of the electric field, weighting potentials and signals.  

The impurity is modeled as a linear gradient between two values measured at either end of each crystal at the time of production.  While the assumption of the the linear profile is something that can be explored, the impurity values at each end of the crystal are well-constrained by measurement.  The depletion voltage is similarly a measured quantity.  However the crystal temperature is somewhat less constrained.  While PT100 temperature sensors within the Quad detector modules provide information of the temperature of the copper cold finger, and the crystal canisters, the temperature in the bulk of the HPGe is assumed, without direct measurement to be uniform through the entire volume at 98~K.  The validity of this assumption can be explored by investigating the temperature dependence of the resolution of pencil beam data sets.   

\subsection{Impact of Temperature on Mobility}
\label{sec:tempDep}

The impact of a change in the assumed temperature of the crystal within the signal generation simulation is to alter the charge carrier mobilities, for both electrons and holes.  The calculated signal shapes are determined by the drift velocity of each charge carrier, which is given by the product of the charge carrier mobility, $\mu$ and the electric field.  In turn $\mu$ is a function of the electric field, the angles of the drift direction and electric field relative to the HPGe crystal axes, and the temperature.  

\begin{figure}[htbp]
    \centering
    \includegraphics[width=\linewidth]{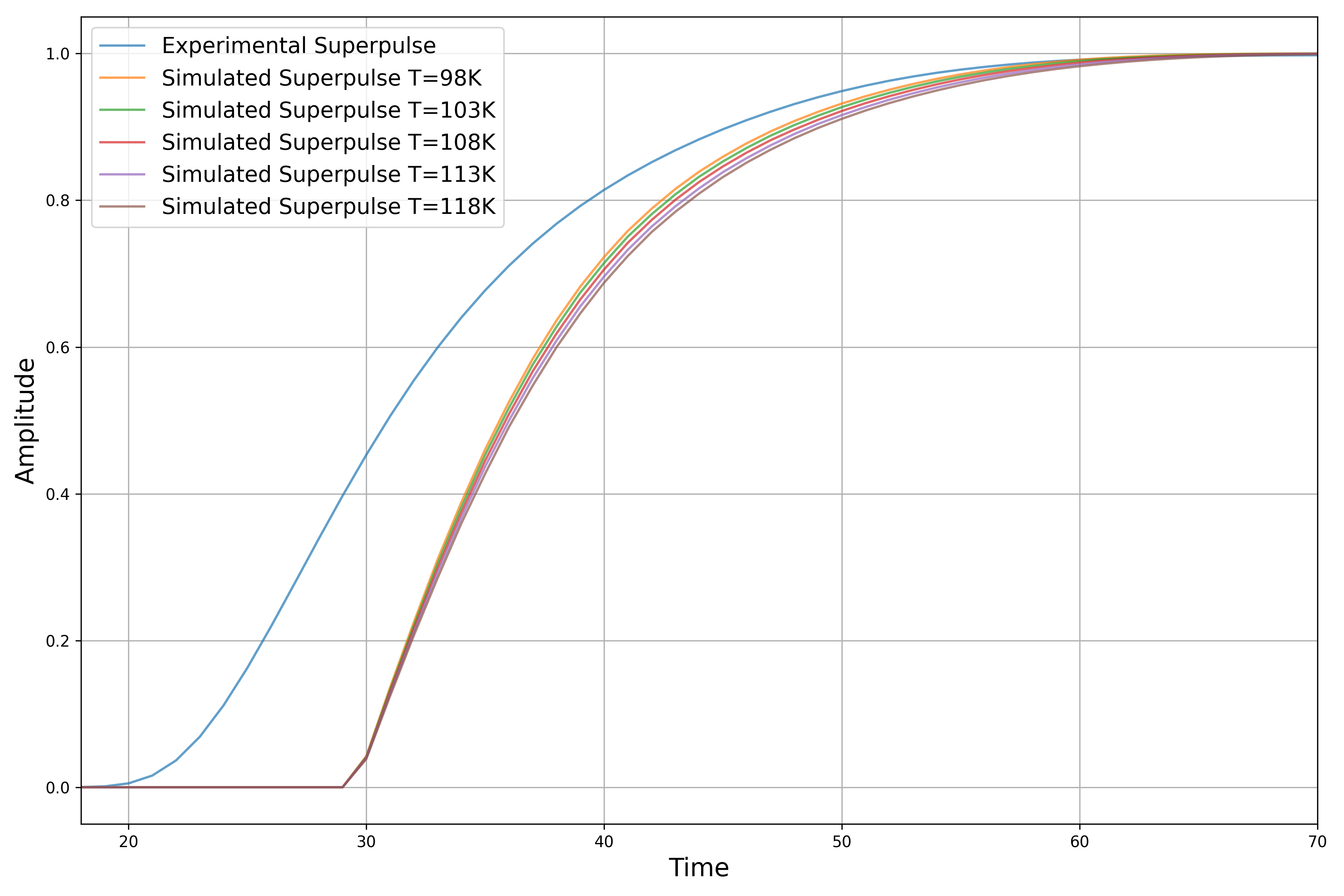}
    \vspace{-1em}
    \caption{Rising edge of experimental signal (blue) as compared to the pristine signal calculated at a range of crystal temperatures.  The sharper rise at lower temperatures is apparent. }
    \label{fig:tempRiseTime}
    \vspace{-1em}
\end{figure}

The temperature dependence of drift velocities, or charge carrier mobilities, has been investigated across a fairly broad range of temperatures~\cite{Omar1987}, and the dependence incorporated into the signal generation code.  In general, the mobility and drift velocities decrease at higher temperatures due to increased scattering with lattice vibration.  The impact of temperature is visible in the pristine signal calculation as shown in Figure~\ref{fig:tempRiseTime}.

However, as it also clear in Fig.~\ref{fig:tempRiseTime}, the change in the signal shape due to temperature is relatively slight in comparison to the impact of the electronics response.  As such, it is of interest to explore the potential impact on the interaction point position resolution, if any.  A similar study~\cite{Prasher2017} exploring the sensitivity of position reconstruction to hole mobility suggested little sensitivity overall, but focused primarily on the reconstructed position.

\subsection{Analysis}
The objective of this analysis is to use the previously described pipeline, including superpulse fitting of the electronics response function using the simplified form, to explore the impact of detector temperature on the accuracy of gamma-ray interaction reconstruction. Specifically, we have varied the temperatures for two GRETA crystals, A35 and A36, both housed in the Q18 Quad Detector module, considering values of $T=98, 103,108, 113, 118$~K.  For each crystal, we have explored the results of the superpulse fitting, in terms of both the mean-squared error of the fit and the parameter values.  

We also used bases produced at each temperature to perform the signal decomposition of a pencil beam in each crystal.  The resulting full-width at half-maximum provides a measure of the position resolution as previously discussed in Section~\ref{subsec:pencilBeamAndResponse}.  

\subsection{Superpulse Fitting Results}

\begin{figure}[!htbp]
    \centering
    \includegraphics[width=\linewidth]{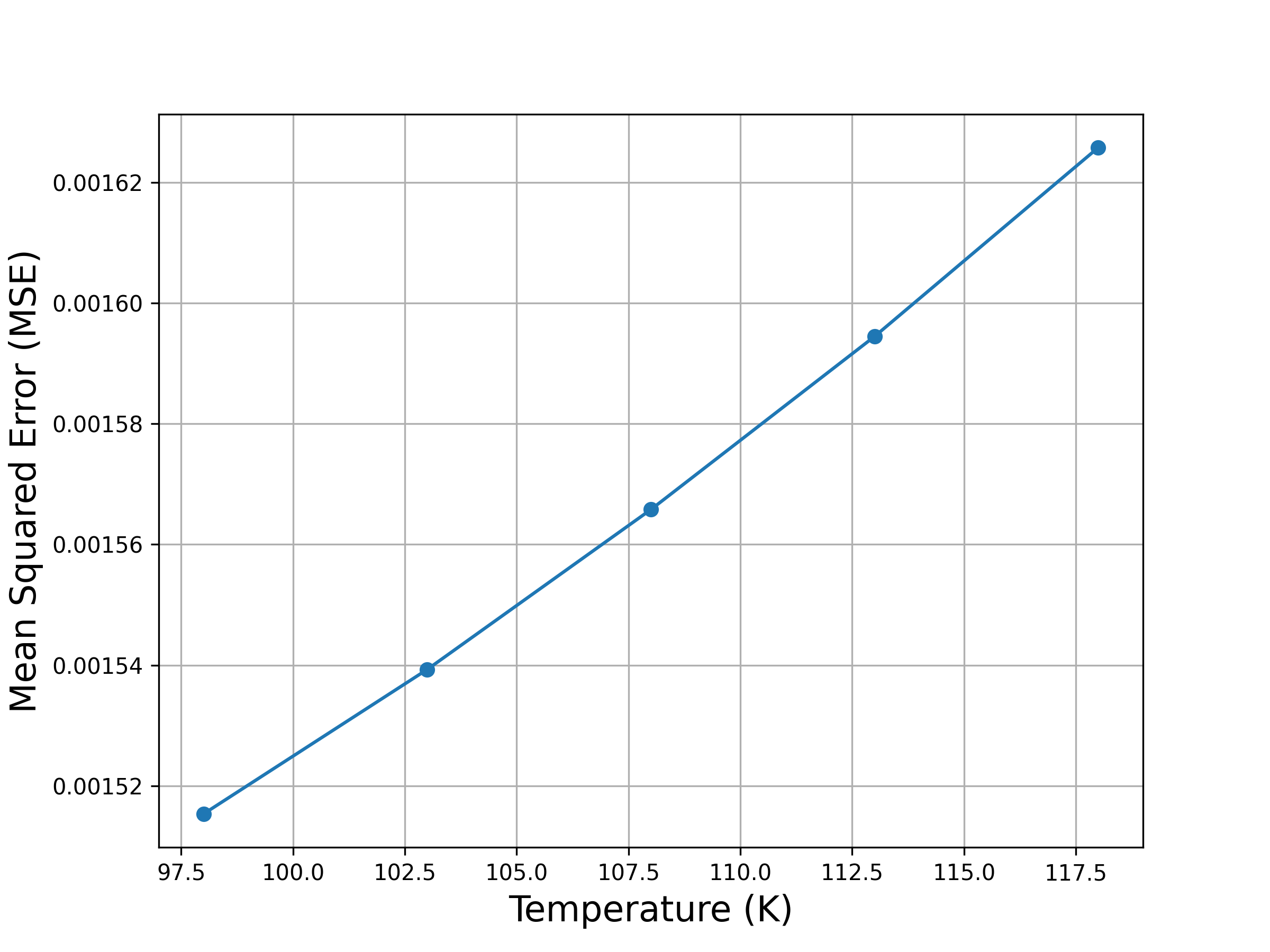}
    \caption{Mean squared error (MSE) plotted versus temperature for the superpulse fitting of crystal A35.  The trend is similar for A36, though the measured MSE are slightly lower for that crystal.}
    \label{fig:mse_vs_temperature_fig}
\end{figure}

We first analyze how varying temperature affects the MSE in the fit of the electronics response model between the experimental and simulated superpulses for each crystal.  The values obtained in the fits for both A35 and A36 are summarized in Table~\ref{tab:mse_vs_temp}, and for A35 are plotted in Fig.~\ref{fig:mse_vs_temperature_fig}.  While there is a clear trend, the maximum variation in MSE is only on the order of 7\%, between the most extreme temperatures 20~K apart.  This suggests that the electronics response parameterization is able to compensate well for the changes in mobility, and the quality of the corrected basis is not strongly impacted.

\begin{table}[!htbp]
\centering
\caption{Mean squared error (MSE) between corrected simulated and measured superpulses across temperatures for both crystals A35 and A36. }
\begin{tabular}{c|c|c}
\textbf{Temperature (K)} & \textbf{MSE for A35} & \textbf{MSE for A36} \\
\hline
98 & {0.001515} & 0.000984\\
103 & 0.001539 & 0.000985\\
108 & 0.001566 & 0.000985\\
113 & 0.001594 & 0.000985\\
118 & 0.001626 & 0.000986\\
\end{tabular}
\label{tab:mse_vs_temp}
\end{table}

\begin{figure} [!htbp]
    \centering
    \includegraphics[width=\linewidth]{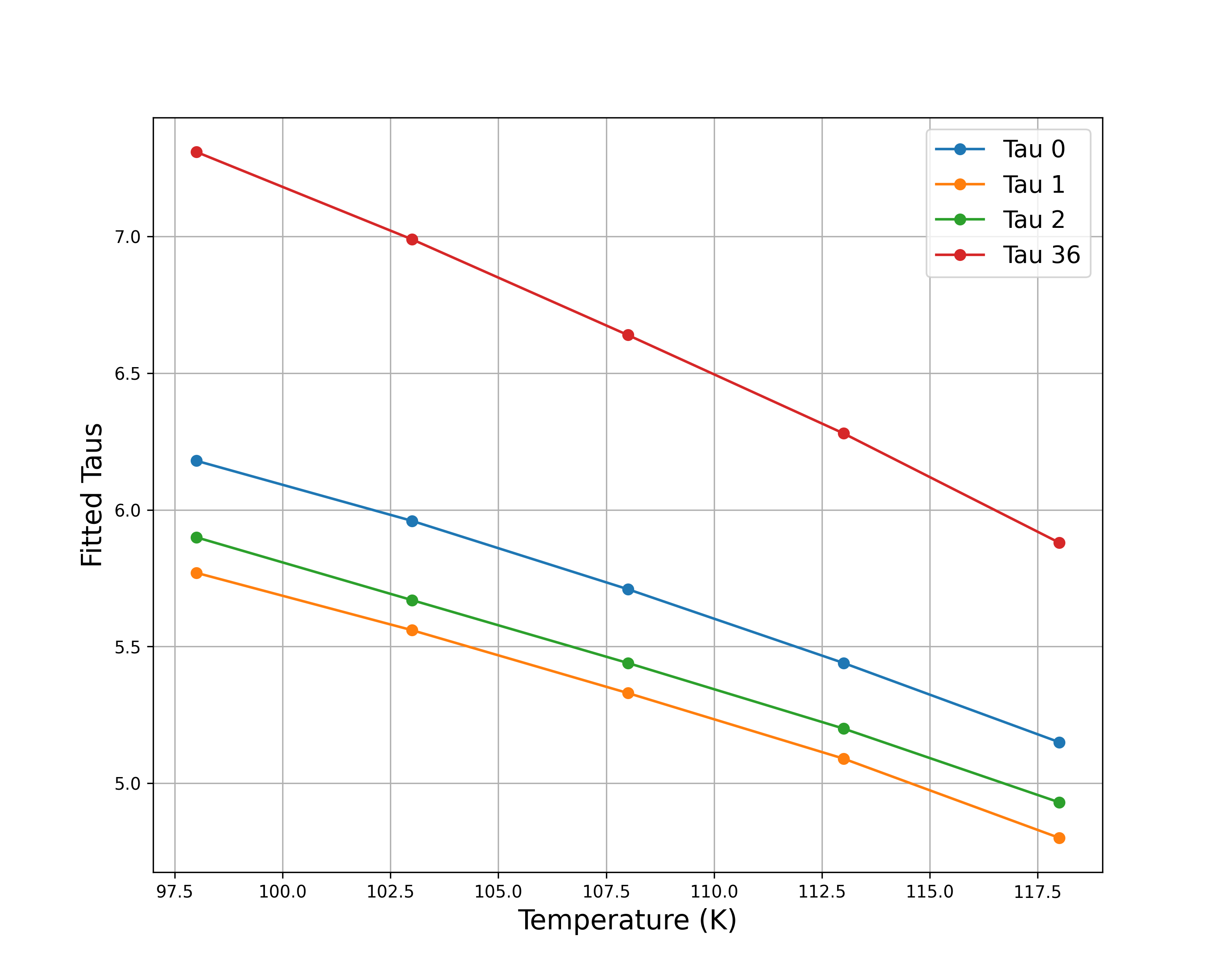}
    \caption{Fitted shaping time constants ($\tau_i$) plotted versus temperature for the superpulse fitting of crystal A35.}
    \label{fig:taus_vs_temperature_fig}
\end{figure}

\begin{figure} [!htbp]
    \centering
    \includegraphics[width=\linewidth]{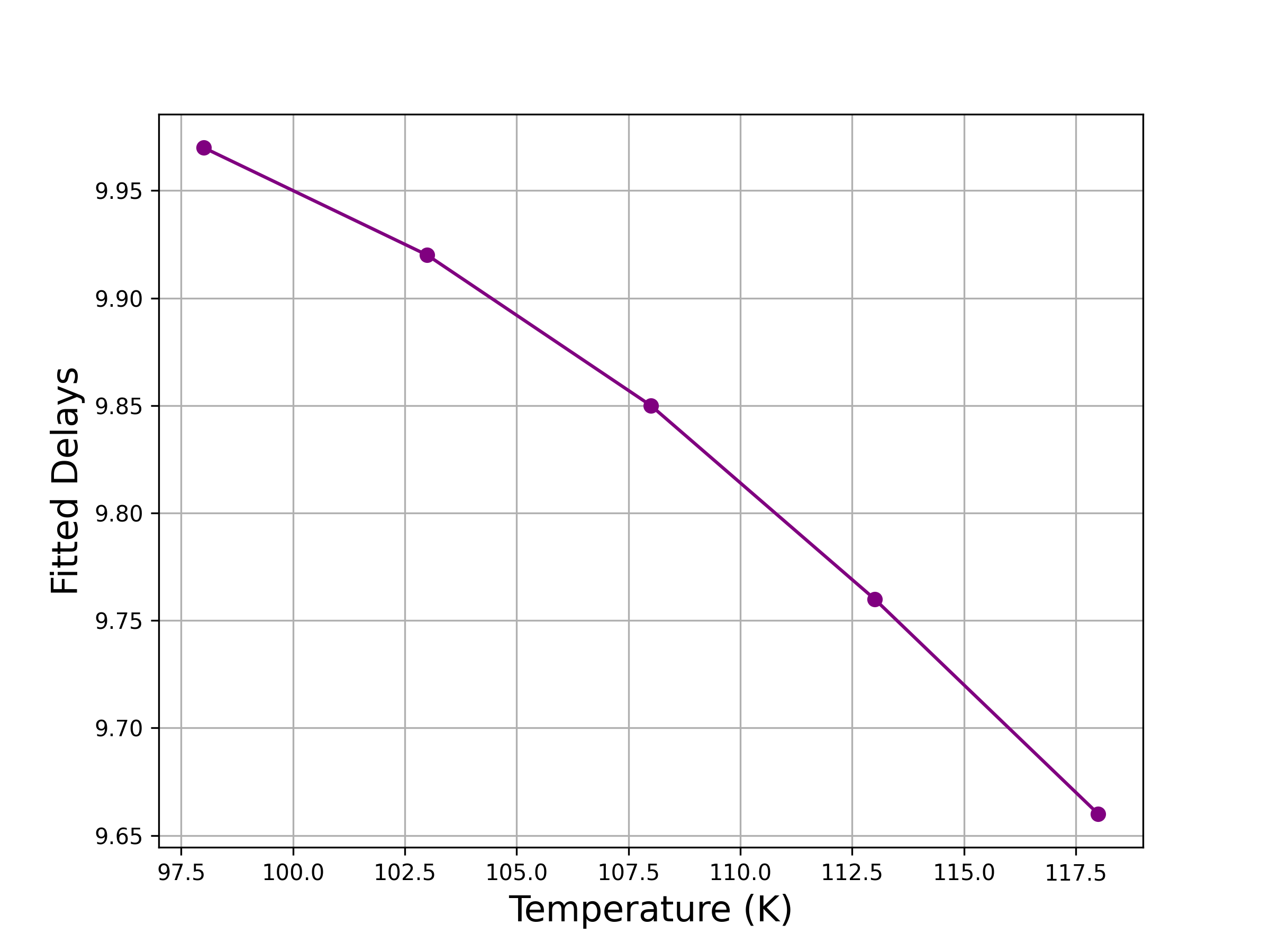}
    \caption{Fitted global time delay plotted versus temperature for the superpulse fitting of crystal A35.}
    \label{fig:delays_vs_temperature_fig}
\end{figure}

We  can also explore the dependence of fitted parameters - we focus here on the results for A35, but the results and interpretation for A36 are very similar.  In Figure~\ref{fig:taus_vs_temperature_fig}, the shaping time constants $\tau_i$ for representative segments $i=0, 1, 2$ and the core-contact (segment 36) are plotted against temperature. Figure \ref{fig:delays_vs_temperature_fig} shows the global time delay for the same temperature range. 

There is a clear trend in parameter value as a function of temperature. The shaping and delay parameters of the electronics response function are sufficient to account for the variation in charge carrier mobility/drift velocity which temperature induces.  As such, it is clear that without additional information to constrain, for example the shaping times, the metrics of the superpulse fit result alone do not allow for an optimization of the temperature in the basis production.

\subsection{Impact on Pencil Beam Position Resolution}
An independent metric for the sensitivity of the basis production to assumed crystal temperature comes analysis of pencil beam data.  We ran the signal decomposition for a pencil beam in crystal A36 and in A36 with our corrected bases produced assuming different temperatures. Our primary metric for position resolution is the numerical FWHM.

%\begin{table}[h]
%\centering
%\caption{FWHM (mm) along X and Y axes for different temperatures (K).}
%\begin{tabularx}{\linewidth}{>{\centering\arraybackslash}Xccc}
%\toprule
%\textbf{Temperature} & \textbf{FWHM\textsubscript{x}} & \textbf{FWHM\textsubscript{y}}\\
%\midrule
%98  & 2.88 & 5.21 \\
%103 & 2.72 & 5.18 \\
%108 & 2.69 & 4.46 \\
%113 & 3.46 & 4.61 \\
%118 & 3.17 & 4.83 \\
%\bottomrule
%\end{tabularx}
%\label{tab:fwhm_results}
%\end{table}

\begin{figure*}[htbp]
    \centering
    \includegraphics[height=0.45\textheight]{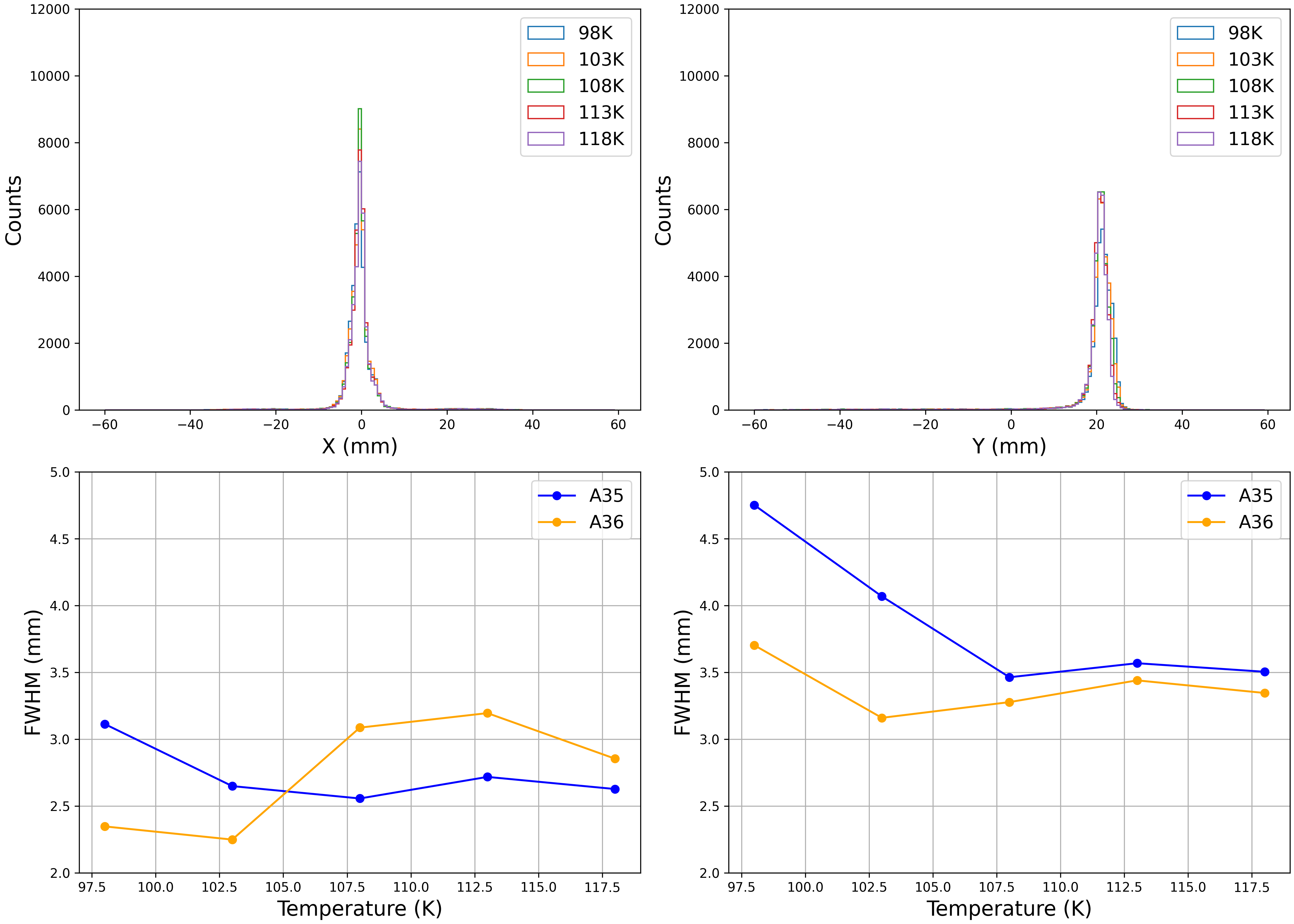}
    \vspace{-1em}
    \caption{Top Panels:  Distributions in x' (left) and y' (right) for the A35 pencil beam decomposed using signal bases produced using different assumed crystal temperatures.  Bottom Panels: FWHM as a function of assumed crystal temperature for the x' (left) and y' (right) distributions of a pencil beam in GRETA crystals A35 and A36.}
    \label{fig:summary_fwhm_vs_temperature_fig}
    \vspace{-1em}
\end{figure*}

The bottom panel in Figure~\ref{fig:summary_fwhm_vs_temperature_fig} summarizes the FWHM as a function of temperature for the x' and y' distributions for both A35 and A36. While there is variation in the numerical FWHM value, this is likely dominated by effects related to the binning of the histograms, for example.  A visual inspection of the distributions for A35 (top panel of Fig.~\ref{fig:summary_fwhm_vs_temperature_fig}) suggests very little impact to the position resolution.  It appears that the electronics response correction dominates the impact of the assumed crystal temperature, and there is little sensitivity in the position reconstruction performance, at least for data sets such as pencil beams. 

\section{Conclusion and Future Directions of Investigation}

The position resolution is a key performance metric for $\gamma$-ray tracking detectors, and is fundamentally linked to the fidelity of the signal basis used to fit experimental signals and extract interaction locations.  The production of signal bases for GRETA$\slash$GRETINA follows an established procedure which has performed well, but exploring approaches to improve and/or simplify the method is of value.  

The simplification of the detector electronic response correction represents such a change, showing comparable performance in the final position resolution while removing 73 degrees of freedom in the superpulse fit.  Next steps related to the electronics response are sensitivity studies of the integral and differential cross-talk parameters, to confirm that the key segment pairs are fully captured. An expansion of the parameter space can also be explored, for example a second shaping time constant to capture higher-order effects in the electronics.

An exploration of the temperature dependence on the position resolution obtained for pencil beams suggests that the superpulse fitting approach is able to compensate for the change in charge carrier mobility that temperature causes.  However, while there is little sensitivity in position resolution with respect to the assumed crystal temperature based on the present work, there are several possible directions for future work to improve basis fidelity, as summarized below.   

\begin{itemize}
    \item \textit{Exploring temperature gradients:} The current pipeline assumes a uniform and constant temperature across the detector crystal. In reality, gradients can develop due to cooling asymmetries and external influences (e.g. proximity to the external environment). While the electronics response correction compensates temperature shifts very effectively, introducing temperature profiles may provide more sensitivity and insight.

    \item \textit{Modifying impurity profiles:} The assumption of a linear impurity profile is one that has not been explored for the case of GRETA$\slash$GRETINA.  Exploring the sensitivity of the performance to the assumption of different profiles may provide important information to improve performance as well as inform our understanding of the crystals themselves.
\end{itemize}

Such explorations are well-supported by the current basis production pipeline, including updates to portions of the code, and could ultimately contribute to improved performance for complex tracking arrays such as GRETA.

This work is supported by the U.S. Department of Energy, Office of Science, Office of Nuclear Physics under contract No. DE-AC02-05CH11231 (LBNL).

\bibliography{citations}

@misc{gretaurl,
  title = {GRETA/GRETINA},
  year = {2021},
  url = {https://greta.lbl.gov},
}

@techreport{riley2020ucgretina,
  author      = {L. A. Riley and D. Weisshaar and H. L. Crawford and M. L. Agiorgousis and C. M. Campbell and M. Cromaz and P. Fallon and A. Gade and S. D. Gregory and E. B. Haldeman and L. R. Jarvis and E. D. Lawson-John and B. Roberts and B. V. Sadler and C. G. Stine},
  title       = {{UCGreta GEANT4 Simulation of the GRETINA Gamma-Ray Energy Tracking Array}},
  institution = {U.S. Department of Energy, Office of Scientific and Technical Information (OSTI)},
  year        = {2020},
  number      = {OSTI ID: 1777581},
  url         = {https://www.osti.gov/servlets/purl/1777581}
}

@ARTICLE{BruyneelADL,
	author = {Bruyneel, B. and Birkenbach, B. and Reiter, P.},
	title = {Pulse shape analysis and position determination in segmented HPGe detectors: The AGATA detector library},
	year = {2016},
	journal = {European Physical Journal A},
	volume = {52},
	number = {3},
	doi = {10.1140/epja/i2016-16070-9},
	type = {Article},
	publication_stage = {Final},
	source = {Scopus},
	note = {Cited by: 50}
}

@article{bentley1975kdtree,
  author    = {Jon Louis Bentley},
  title     = {Multidimensional Binary Search Trees Used for Associative Searching},
  journal   = {Communications of the ACM},
  volume    = {18},
  number    = {9},
  pages     = {509--517},
  year      = {1975},
  publisher = {ACM},
  doi       = {10.1145/361002.361007},
  note      = {1975 ACM Student Award Paper: Second Place}
}

@article{DirkGRETINAPerformance,
title = {The performance of the gamma-ray tracking array GRETINA for gamma-ray spectroscopy with fast beams of rare isotopes},
journal = {Nuclear Instruments and Methods in Physics Research Section A: Accelerators, Spectrometers, Detectors and Associated Equipment},
volume = {847},
pages = {187-198},
year = {2017},
issn = {0168-9002},
doi = {https://doi.org/10.1016/j.nima.2016.12.001},
url = {https://www.sciencedirect.com/science/article/pii/S0168900216312402},
author = {D. Weisshaar and D. Bazin and P.C. Bender and C.M. Campbell and F. Recchia and V. Bader and T. Baugher and J. Belarge and M.P. Carpenter and H.L. Crawford and M. Cromaz and B. Elman and P. Fallon and A. Forney and A. Gade and J. Harker and N. Kobayashi and C. Langer and T. Lauritsen and I.Y. Lee and A. Lemasson and B. Longfellow and E. Lunderberg and A.O. Macchiavelli and K. Miki and S. Momiyama and S. Noji and D.C. Radford and M. Scott and J. Sethi and S.R. Stroberg and C. Sullivan and R. Titus and A. Wiens and S. Williams and K. Wimmer and S. Zhu}
}

@article{GRETINAPaschalis,
title = {The performance of the Gamma-Ray Energy Tracking In-beam Nuclear Array GRETINA},
journal = {Nuclear Instruments and Methods in Physics Research Section A: Accelerators, Spectrometers, Detectors and Associated Equipment},
volume = {709},
pages = {44-55},
year = {2013},
issn = {0168-9002},
doi = {https://doi.org/10.1016/j.nima.2013.01.009},
url = {https://www.sciencedirect.com/science/article/pii/S0168900213000508},
author = {S. Paschalis and I.Y. Lee and A.O. Macchiavelli and C.M. Campbell and M. Cromaz and S. Gros and J. Pavan and J. Qian and R.M. Clark and H.L. Crawford and D. Doering and P. Fallon and C. Lionberger and T. Loew and M. Petri and T. Stezelberger and S. Zimmermann and D.C. Radford and K. Lagergren and D. Weisshaar and R. Winkler and T. Glasmacher and J.T. Anderson and C.W. Beausang}
}

@article{Prasher2017,
title = {Sensitivity of GRETINA position resolution to hole mobility},
journal = {Nuclear Instruments and Methods in Physics Research Section A: Accelerators, Spectrometers, Detectors and Associated Equipment},
volume = {846},
pages = {50-55},
year = {2017},
issn = {0168-9002},
doi = {https://doi.org/10.1016/j.nima.2016.11.038},
url = {https://www.sciencedirect.com/science/article/pii/S0168900216311925},
author = {V.S. Prasher and M. Cromaz and E. Merchan and P. Chowdhury and H.L. Crawford and C.J. Lister and C.M. Campbell and I.Y. Lee and A.O. Macchiavelli and D.C. Radford and A. Wiens}
}

@misc{trf,
    author = {Tung-Yen Wang},
    title = {Trust-region methods},
    url = {https://optimization.cbe.cornell.edu/index.php?title=Trust-region_methods},
    year = {accessed on December 2025}
}

@misc{radford2018,
  author       = {David Radford},
  title        = {{GRETINA / GRETA Signal Decomposition}},
  howpublished = {AGATA-GRETA Meeting, CSNSM, April 2018},
  institution  = {Oak Ridge National Laboratory},
  year         = {2018},
  url          = {https://indico.in2p3.fr/event/16944/contributions/60379/attachments/47864/60286/Radford_decomp_Apr2018.pdf}
}

@article{Schockley,
	author={W. Shockley},
	title = {"Currents to Conductors Induced by a Moving Point Charge}, 
	journal={Journal of Applied Physics},
	volume = {9}, 
	page = {635}, 
	year = {1938},
	doi={https://doi.org/10.1063/1.1710367}
}

@article{Ramo,
	author = {S. Ramo},
	title = {Currents Induced by Electron Motion},
	journal = {Proceedings of the IRE},
	volume = {27}, 
	number = {9}, 
	page = {584},
	year = {1939},
	doi = {10.1109/JRPROC.1939.228757}
}

@misc{cromaz2024signaldecomp,
  author       = {Mario Cromaz},
  title        = {Signal Decomposition AI/ML Progress},
  howpublished = {Presentation at the 4th AGATA-GRETINA/GRETA Collaboration Meeting},
  year         = {2024},
  month        = nov,
  organization = {Lawrence Berkeley National Laboratory},
  note         = {Available at \url{https://indico.in2p3.fr/event/32594/contributions/146474/attachments/89594/135703/SignalDecompML.pdf}},
}

@InProceedings{ pandas,
  author    = { {W}es {M}c{K}inney },
  title     = { {D}ata {S}tructures for {S}tatistical {C}omputing in {P}ython },
  booktitle = { {P}roceedings of the 9th {P}ython in {S}cience {C}onference },
  pages     = { 56 - 61 },
  year      = { 2010 },
  editor    = { {S}t\'efan van der {W}alt and {J}arrod {M}illman },
  doi       = { 10.25080/Majora-92bf1922-00a }
}

@article{Omar1987,
title = {Drift velocity and diffusivity of hot carriers in germanium: Model calculations},
journal = {Solid-State Electronics},
volume = {30},
number = {12},
pages = {1351-1354},
year = {1987},
issn = {0038-1101},
doi = {https://doi.org/10.1016/0038-1101(87)90063-3},
url = {https://www.sciencedirect.com/science/article/pii/0038110187900633},
author = {M.Ali Omar and Lino Reggiani}
}

@article{bevington,
    author = {Bevington, Philip R. and Robinson, D. Keith and Bunce, Gerry},
    title = {Data Reduction and Error Analysis for the Physical Sciences, 2nd ed.},
    journal = {American Journal of Physics},
    volume = {61},
    number = {8},
    pages = {766-767},
    year = {1993},
    month = {08},
    issn = {0002-9505},
    doi = {10.1119/1.17439},
    url = {https://doi.org/10.1119/1.17439},
    eprint = {https://pubs.aip.org/aapt/ajp/article-pdf/61/8/766/12168797/766_2_online.pdf},
}
\end{document}